\documentclass{aastex63}

\def\a{\"a}
\def\o{\"o}
\def\u{\"u}
\def\A{\"A}
\def\O{\"O}
\def\U{\"U}

\def\gapprox{\lower.4ex\hbox{$\;\buildrel >\over{\scriptstyle\sim}\;$}}
\def\lapprox{\lower.4ex\hbox{$\;\buildrel <\over{\scriptstyle\sim}\;$}}
\def\ref#1{\par\noindent\hangindent1cm {#1}}

\begin{document}

\title{Finite System-Size Effects in Self-Organized Criticality Systems}

\correspondingauthor{Markus J. Aschwanden}
\email{aschwanden@lmsal.com}

\author{Markus J. Aschwanden}

\affiliation{Solar and Stellar Astrophysics Laboratory (LMSAL),
 Palo Alto, CA 94304, USA}

\begin{abstract}
We explore upper limits for the largest avalanches or catastrophes 
in nonlinear energy dissipation systems governed by {\sl self-organized 
criticality (SOC)}. We generalize the idealized ``straight'' power low 
size distribution and Pareto distribution functions in order to 
accomodate for incomplete sampling, limited instrumental sensitivity, 
finite system-size effects, ``Black-Swan'' and ``Dragon-King'' 
extreme events. 
Our findings are: (i) Solar flares show no finite system-size limits
up to $L \lapprox 200$ Mm,
but solar flare durations reveal an upper flare duration limit of
$\lapprox 6$ hrs; (ii) Stellar flares observed with KEPLER exhibit
inertial ranges of $E \approx 10^{34}-10^{37}$ erg, finite system-size
ranges at $E \approx 10^{37}-10^{38}$ erg, and extreme events at
$E \approx (1-5) \times 10^{38}$ erg; (iii) The maximum flare energy
of different spectral-type stars (M, K, G, F, A, Giants)
reveal a positive correlation with the stellar radius, which indicates a 
finite system-size limit imposed by the stellar surface area. Fitting our 
finite system-size models to terrestrial data sets (Earth quakes,
wildfires, city sizes, blackouts, terrorism, words, surnames,
web-links) yields evidence (in half of the cases) for finite system-size
limits and extreme events, which can be modeled with dual power law 
size distributions.
\end{abstract}


\section{	Introduction			}

At the time of writing this paper, while the author works
in smoke-filled poor air, caused by 
large wildfires around San Francisco, {\sl The New York Times}
(September 2000) reported that the number of observed
wildfires in California doubled 
in number as well as in size this year, most likely caused by
the global warming that contributes to drier conditions
of forests and thus to their enhanced flammability. 
An obvious question arises 
therefore, to what maximum size the largest possible wildfire 
could grow. The size of wildfires is generally measured 
by the burned area (in units of acres). In the USA, of the 
1.4 million wildfires that have occurred since 2000, 197 exceeded 
100,000 acres, and 13 exceeded 500,000 acres, according to
a report of the {\sl Interagency Fire Center (NIFC)},
(https://fas.org/sgp/crs/misc/IF10244.pdf).

Although wildfires represent just one phenomenon of
nonlinear energy dissipation events that have been modeled
with {\sl self-organized criticality (SOC)} 
(Drossel and Schwabl 1992; Malamud et al. 1998; Zinck and Grimm 2008; 
Hergarten 2013), a fundamental
question is the determination of finite system-size effects,
which constrains the largest possible event in a size
distribution. Here we investigate finite system-size effects
in large astrophysical data sets (with $\gapprox 10^3$ events),
which mostly apply to solar and stellar flare statistics
(with 16 data sets). In the spirit of interdisciplinary research, 
we subject also 8 empirical data sets to our analysis,
selected by Clauset et al.~(2009), covering information technology
(words, surnames, weblinks), geophysics 
(Earth quakes, wildfires), and human activities
(city sizes, blackouts, terrorism statistics).

The SOC concept predicts power law-like size distributions of
nonlinear energy dissipation events (also called avalanches
or catastrophes), as well as scaling laws in form of
power law relationships (Aschwanden 2020).
However, although power laws are the
hallmark of SOC avalanches, many observed size distributions
exhibit significant deviations from ideal power law functions,
caused by critical thresholds, instrumental sensitivity limitations,
background subtraction, incomplete sampling, finite system-size
effects, and extreme event outliers. In this study we demonstrate
that all these effects can be built into generalized power law
models that fit the data within acceptable values of least-square 
fits ($\chi \lapprox 2$). Alternative size distribution functions
have also been tested, such as a Poisson, log-normal, exponential,
stretched exponential, or power law plus cutoff (Clauset et al.~2009).
The modeled size distributions
may show large parameter ranges with ideal power law behavior
for some size distributions on one side, but also curved power laws 
without any straight power law segments in other cases, which triggered 
some researchers to question whether power law size distribution 
functions exist at all (Stumpf and Porter 2012). Nevertheless,
we find that straight power laws still exist for large inertial ranges 
(of $\approx 2-5$ decades), but are strongly convolved 
in data sets with small inertial ranges (of $\approx 1-2$
decades). The promise of this study is the detectability of
finite system sizes in form of exponential-like cutoffs 
near the upper end of an observed size distribution,
while a straight power law all the way to the largest 
observed event indicates a lack of finite system-size
effects, and thus even larger events are expected when
the statistics is prolonged in time. In other words,
we attempt to discriminate between a temporary maximum size
(for short-term data sets) and an absolute maximum size 
(for long-term data sets). And there is also the notion
of ``Dragon-King'' and ``Black-Swan'' events, 
which represent outliers from a
canonical power law size distribution, and thus may be
produced by an alternative physical mechanism altogether
(Sornette 2009; Sornette and Ouillon 2012, measured
for solar and stellar flares for the first time in 
Aschwanden (2019). 
The Black-Swan theory is essentially a metaphor for rare 
events that cannot be predicted by standard statistics
(Taleb 2007). Dragon-King events, defined by Sornette
2009 in financial risk management, {\sl are fundamentally 
the same: an unusually extreme event that was unforeseen 
by almost everyone, even when examining the far right tail 
of the loss distribution.} For astrophysical applications,
Black-Swan or Dragon-King events are produced by different
physical mechanisms than events in the left tail of power law
distributions. The precise model for
such outliers is of course unknown, but we identify them
by an excess in the size distribution above the 
exponential-like cutoff of the finite system-size limit.
{\sl It has been shown that the solar flare events  
observed in the past four centuries have not 
exceeded the level of the largest flares 
observed in the space era, and that there is at most 
a 10\% chance of a flare larger than about X30 
(GOES-class) in the next 30 years} (Schrijver et al.~2012).

The content of this paper contains theoretical definitions
of power law-like size distributions (Section 2), data
analysis of 24 astrophysical and empirical data sets with
discussion (Section 3), and conclusions (Section 4). 

\section{	Theory and Method 				}

Our goal is the determination of the power law slope $\alpha_x$
in power law-like size distributions, by modeling a number of
known effects, such as the (astrophysical) background subtraction
(Section 2.1), incomplete sampling, instrumental sensitivity
limitations, and detection thresholds, which affect 
differential size distributions (Secions 2.2) and require thresholded
power law distributions (Section 2.3), finite system-size effects 
(Section 2.4), and extreme (``Dragon-King'') events (Section 2.5).
The least-square optimization criterion is described in Section
2.6.

\subsection{    Empirical Background Subtraction        }

We start with a data sample $X_i, i=1,...,n_{ev}$ that contains
some size parameter $X_i$ with $n_{ev}$ events. In astrophysical 
data sets, observables for the size parameter $X_i$ are most 
often measured in terms of peak fluxes $P_i$, time-integrated 
fluences, intensities, luminosities, or energies $E_i$, and event
time durations $T_i$. The observable $X_i$ refers to the size
of an event, a pulse amplitude, or a flare. A first correction that needs
to be applied to the size value $X_i$ is the background subtraction
\begin{equation}
	x_i = X_i - B_i \ , \quad i=1,...,n_{ev} \ ,
\end{equation}
where $B_i$ represents an event-unrelated flux or fluence,
for instance the galactic quiescent soft X-ray flux that is
irradiated from the same source direction as the 
soft X-ray flux of a solar flare. If the flux time profile $F_i(t)$
of an event ``i'' is available, the background flux can be
determined from the pre-flare flux $B_i$, and the corrected
peak flux is $P_i = max[F_i(t)] - B_i$. Since the background
correction affects the power law slope $\alpha_P$ of the
peak flux (see Fig.~5 in Aschwanden 2011), it is advisable
to apply a mean estimated background correction $B=<B_i>$ 
at least. 

\subsection{    Differential Size Distribution          }

A {\sl size distribution} of events, $N(x) dx$, also called
a {\sl differential occurrence frequency distribution},
can ideally be approximated by a power law function
as a function of some size parameter $x$, quantified
by three observables ($x_1, x_2, n_0$) and one variable ($\alpha_x$),
\begin{equation}
        N(x) dx = n_0 \ x^{-\alpha_x} dx \ ,
        \qquad x_1 \le x \le x_2 \ ,
\end{equation}
where $x_1$ and $x_2$ are the lower and upper bounds of
the power law inertial range, 
$n_0$ is a normalization constant, and  
$\alpha_x$ is the power law slope. 
Uncertainties
in previous calculations of the power law slope mostly resulted from
the arbitrary choice of fitting ranges $[x_1, x_2]$, which
can be largely corrected by a generalized power law function
that includes under-sampling at the lower end and finite
system-size effects at the upper end of the inertial range
$[x_1, x_2]$.

\subsection{    Thresholded Power Law Distribution              }

The straight power law function (Eq.~2) can be generalized
with a threshold parameter $x_0$. This generalized function is
called {\sl Lomax distribution} (Lomax 1954),
{\sl the generalized Pareto distribution} (Hosking and Wallis
1987), or the {\sl Thresholded power law size distribution}
(Aschwanden 2015),
\begin{equation}
        N(x) dx = n_0 \left( x_0 + x \right)^{-\alpha_x} dx \ ,
\end{equation}
where $n_0$ is a normalization constant.
This threshold parameter $x_0$ models three different
features: such as truncation effects due to incomplete sampling of
events below a threshold (if $x_0 > 0$), incomplete
sampling due to instrumental sensitivity limitations (if $x_0 > 0$),
or subtraction of event-unrelated background (if $x_0 < 0$),
as it is common in astrophysical data sets (Aschwanden 2015, 2019).

\subsection{    Finite System-Size Effects              }

The size distribution can be uniquely approximated with a classical
power law function, if the lower bound ($x_1$) and upper bound
($x_2$) is well-defined. In practice, however, the lower bound
is flattened by undersampling, detection thresholds, or instrumental
sensitivity limitations, while the upper bound typically shows a gradual
steepening due to finite system-size effects (Pruessner 2012).
Ignoring these effects
leads to power law fits with arbitrary inertial ranges $[x_1, x_2]$,
which affects the accuracy of the determined power law slope and
its uncertainty. Finite system size effects are modeled here with
an exponential cutoff function, which we can
combine with the generalized Pareto distribution (Eq.~3),
\begin{equation}
        N(x) dx = n_0 \left( x_0 + x \right)^{-\alpha_x}
        \exp{ \left( - {x \over x_e}\right) } \ dx \ ,
\end{equation}
which is quantified with an exponential function
$\exp(-x/x_e)$ at a characteristic size $x \approx x_e$. 
The inclusion of finite
system-size effects steepens the power law slope asymptotically to
infinity at the upper end of the distribution function. 

\subsection{    Extreme Events  }

Sometimes extreme events at the upper end of the size 
distribution function cannot be fitted with the exponential 
cutoff function, as expected from finite system-size effects.

These extreme events that deviate
from a standard power law distribution function have also been
dubbed as ``Black-Swan'' and {\sl ``Dragon-King'' events} 
(alluding to their
extremely rare appearance) by Sornette (2009)
and Sornette and Ouillon (2012), who
suggested that they are generated by a different physical
mechanism. Detections of such extreme event outliers have
occasionally  been noted in astrophysical data sets
(Aschwanden 2019).

To accomodate these extreme events, we define a size distribution
with a second power law component, where the first power law
distribution includes the exponential function (Eq.~4) with
amplitude $(1-q_{pow})$, while the second power law distribution
with amplitude $(q_{pow})$ extends without a cutoff
all the way to the largest event (with an identical power law 
slope, in order to minimize the number of free parameters),
\begin{equation}
        N(x) dx = n_0 \left( x_0 + x \right)^{-\alpha_x}
        \left[(1-q_{pow}) \exp{ \left( - {x \over x_e}\right) }
        + q_{pow} \right] \ dx \ ,
\end{equation}
where we define the exponential cutoff energy $x_e=x_2\ q_{exp}$,
with $q_{exp}$ being a free parameter.
This combinatory definition of the size distribution converges
to the canonical finite system-size distribution as defined
in Eq.~(4) for $q_{pow} \mapsto 0$, and to the previously
defined Pareto distribution function for $q_{pow} \mapsto 1$. 

A synopsis of the three models is shown in Fig.~1, depicting
the Pareto distribution model (PM), the finite system-size
effect model (FM), and combined with the extreme event component 
model (EM). Note that the most general model (EM) has 3 
observables ($n_0, x_0, x_2$) and 3 variables 
($\alpha_x, q_{exp}, q_{pow}$).

\subsection{	Fitting Method		}

Our power law fitting procedure is straightforward. As indicated
in Fig.~1, the total range of the size distribution is bound by
the smallest ($x_1 = min[x_i]$) and the largest event size
($x_2 = max[x_i]$). Before any data binning is done, we apply
first the empirical background subtraction (Section 2.1). 
Then we bin the number or counts of events 
per bin, $C_j = N_j \Delta x_j$, with logarithmic steps $\Delta x_j$,
and determine the threshold $x_0$ from the maximum of the 
binned count rate, $max[C_i] = C_i(x_0)$. The size $x_0$ represents
a threshold in the Pareto distribution that roughly demarcates
the range $[x_1, x_0]$ of incomplete sampling, while the range 
$[x_0, x_2]$ represents the {\sl inertial range} where the 
size distribution is fitted. The lower bound $x_0$ of the
inertial range is adjusted when multiple peaks are present 
(see also Monte-Carlo simulations in Aschwanden 2011; 2015).

The fitting of any of the 3 models of the differential 
occurrence size distribution $N_{diff}^{theo}(x)$ to an 
observed (binned) size distribution $N_{diff}^{obs}(x)$
is performed with a standard least-square $\chi^2$-criterion
(i.e., reduced $\chi^2$),
\begin{equation}
	\chi_{diff} =
	\sqrt{ {1 \over (n_{bin} - n_{par})}
	\sum_{j=1}^{n_{bin}}
	{[N_{diff}^{theo}(x_j)-N_{diff}^{obs}(x_j)]^2
	\over \sigma_{diff,j}^2 }
	} \ , 
\end{equation}
where $x_j, j=1,...,n_j$ are the counts per bin width,
$n_j$ is the number of the bins,
and $n_{par}$ is the number (1, 2, 3) of free parameters
(variables) of the fitted model functions.
For the number of bins we use 6 bins per decade.
The estimated uncertainty of counts per bin,
$\sigma_{diff,j}$ (Eq.~6), is according to Poisson statistics,
\begin{equation}
        \sigma_{diff,j} = {\sqrt{N_j \Delta x_j} \over \Delta x_j} \ .
\end{equation}
where $\Delta x_j$ is the (logarithmic) bin width.
The goodness-of-fit $\chi_{diff}$ quantifies which
model size distribution is consistent with the (observed) data.

Finally, the uncertainty $\sigma_\alpha$ of the best-fit power law
slope $\alpha$ is estimated to,
\begin{equation}
        \sigma_\alpha = {\alpha \over \sqrt{n_{ev}}} \ ,
\end{equation}
with $n_{ev}$ the total number of events in the entire size 
distribution (or in the fitted range), according to 
Monte-Carlo simulations with least-square fitting 
(Aschwanden 2011). A slightly different estimate of
$\sigma_\alpha=(\alpha-1)/\sqrt{n}$ is calculated in 
Clauset et al.~(2009), {\sl based on the Hill estimator
(Hill 1975) for the maximum likelihood estimator of the
scaling parameter.}  

\section{	Data Analysis and Discussion	}

Theoretical models of SOC systems often involve a
large 1-D, 2-D, or 3-D lattice grid
that has a finite system size. As long as avalanches
evolve inside the lattice grid, we expect a power
law distribution of avalanche sizes, while
avalanches propagating to the edge of the lattice
grid have a quenched or reduced
size that manifests itself as an exponential-like
cutoff in the size distribution (Pruessner 2012).
The presence of this suppression
effect enables a measurement of the system size.
Such a measurement, however, requires
sufficiently long sampling times, otherwise one can
not distinguish between a power law with a sharp
cutoff versus an exponential cutoff.
Fortunately, the data set of stellar
flares observed with KEPLER provides sufficient
statistics to infer two power law components in
the size distribution. In the following we
investigate the presence or absence of
finite system-size limits in solar and stellar
flare data.

\subsection{	Performance of the 3 Models 	}

Ultimately, the aim of this analysis is adequate modeling of the 
observed power law size distributions, to which we apply three
different theoretical models; (i) the Pareto distribution model
that shows
a flattening at the lower end due to incomplete sampling, with a
straight power law at the upper end (PM); (ii) the Pareto
distribution model with an exponentially dropping cutoff at the
upper end that is caused by finite system-size effects (FM);
and (iii) an additional power law component that extends above
the exponential cutoff, caused by extreme events that are also 
called ``Dragon-King'' extreme event model (EM). These three 
models are depicted in Fig.~1. The three models have an increasing 
number of free parameters (1, 2, and 3), where the models with 
a larger number of free parameters
represent generalizations of the models with fewer parameters.
The Pareto distribution, the simplest of our three size distribution models,
has three observable parameters: 
the threshold $x_0$, the maximum value $x_2$, 
the normalization constant $n_0$ (Eq.~3),
and one variable: the power law slope $\alpha_x$.
The finite system-size model contains an additional size parameter 
$x_e = q_{exp}\ x_2$ that marks the size of the exponential cutoff
(Eq.~4). The extreme event model contains an additional parameter
$q_{pow}$ that expresses the fraction of extreme events (Eq.~5).

The best-fitting cases of the 24 analyzed data sets with a 
goodness-of-fit $\chi \lapprox 2$ (fitted over the 
inertial range) are listed in Table 1, containing the number of
events $n_{ev}$ per data set, the power law slope $\alpha_x$, the
goodness-of-fit $\chi$, and the best-fit model (PM, FM, EM).
According to this evaluation listed in Table 1 we find that 
6 data sets are fitted best with the Pareto model (PM), 
4 data sets are fitted best with the finite system-size model 
(FM), and 8 data sets are
fitted best with the extreme event model (EM). 
Thus each of the 3 models 
has its own merit to fit real-world data. In the following we analyze 
finite system-size effects and deviations from ideal power law
distributions.

\subsection{	Solar Flare Finite Size Limit		}

The first six analyzed data sets were compiled from some
$n_{ev} \lapprox 10^4$ solar flares, detected with three different
instruments: the {\sl Hard X-ray Burst Spectrometer
(HXRBS)} onboard the {\sl Solar Maximum Mission (SMM)},
the {\sl Burst And Transient Source Experiment (BATSE)}
onboard the {\sl Compton Gamma Ray Observatory (CGRO)}, and 
the {\sl Ramaty High-Energy Solar Spectroscopic Imager
(RHESSI)}, where each instrument was operated during a
different solar cycle (Aschwanden 2015; 2019; 
and references therein).
The first three data sets were measured from the 
flare peak count rate $P$ (Figs.~2a, 2b, 2c), while the
next three sets were measured from the time-integrated
counts (or fluences) $F$ of hard X-ray photons with 
energies of $E$ $\gapprox 20$ keV (Figs.~2d, 2e, 2f). 
Each histogram shown in Fig.~2 contains in the uppermost
bin the absolute maximum size $x_2$ of the largest event.
The fact that the size distributions of these six 
data sets observed in solar flares do not reveal any
recognizable finite system-size limit, leaves the
progression of the power law distribution at the upper
end open, implying that larger ``super events'' are 
possible, beyond the observed maximum event size $x_2$. 
This brings us to the question what are the size limits 
in solar flares? One of the largest observed 
spatial scale of a solar flare (or active region)
was measured during the ``Bastille-day'' flare,
amounting to about $L \approx 200$ Mm $\approx 0.3$ 
solar radius (Aschwanden and Alexander 2001).
Interestingly, the depth of the solar convection zone 
has a similar spatial scale in vertical direction, 
above a radius of $R \approx 0.7 R_{\odot}$. 
This coincidence between the maximum size of the
horizontal flare area and the vertical depth of the 
convection zone may play a role in predicting the depths
of stellar convection zones. The finding of a
correlation between the stellar radius and flare energy
as shown in Fig.~7 is an encouraging result along this
line.

\subsection{	Solar Flare Finite Duration Limit 	}

Size distribution fits of solar flare durations $T$ are shown
in Figs.~3a, 3b, and 3c. All three cases can be fitted with
an exponential cutoff over an inertial range of $\lapprox 2$
decades, covering an inertial range of $T \approx 10^2-10^4$ s,
or from $\approx 2$ minutes to 3 hours. The inferred power law
slopes have a mean of $\alpha_T = 1.90\pm0.09$ (i.e., the
average of the 3 cases shown in Fig.~3), which is close to
the prediction of SOC models ($\alpha_T = 2.0$; Aschwanden and
Freeland 2012). The meaning of a finite system size in the
time domain of this data set of flare durations $T$ is a temporal
limit that cannot be exceeded by a coherent flare process. 
We see that the longest flare durations are
$T_2 \approx 5.6$ hrs for HXRBS data (Fig.~3a),
$T_2 \approx 2.2$ hrs for BATSE data (Fig.~3b), and
$T_2 \approx 0.8$ hrs for RHESSI data (Fig.~3c).
Since each data set is recorded by a different instrument
and during different (non-overlapping) time epochs,
the differences in the maximum event duration $T_2$ could
result from different definitions of the flare duration
or detection threshold.
Nevertheless, all three data sets seem to indicate an exponential
cutoff, which implies an upper limit of the flare duration,
in the order of $T_2 \lapprox 6$ hrs. The exponential
cutoff suggests that we
are unlikely to observe solar flares with longer durations.
We could not make such a statement if the size distribution 
of time durations would show a straight power law all the
way to the upper end.

We may ask why flare durations (Fig.~3) do not show
power laws extending over 3-5 decades as we find for
peak fluxes and fluences (Fig.~2). We note that
the inertial range with straight power law behavior 
is only 1-2 decades in size distributions of event 
durations, which makes it harder to separate it from
the flattening at the lower end ($x \lapprox x_0$) and the
exponential cutoff at the upper end ($x \gapprox x_3$).
However, at this time it is not entirely understood why 
event durations have a pronounced exponential cutoff (Fig.~3),
in contrast to event peak rates and fluences (Fig.~2).
It appears that finite system-size effects are different 
in the time domain (e.g., flare durations) and in the
spatial domain (see Fig.~10 and Appendix B).

\subsection{	Stellar Flare Finite Energy Limits     }

The size distribution of stellar flares according to the 
entire KEPLER (Borucki et al.~2010)
flare event catalog (Davenport 2016; Yang and Liu 2019)
includes a total of 
162,264 flare events from different stars (normalized 
to the same stellar distance) and is shown in Fig.~(4).
This size distribution describes the largest available 
solar and stellar flare data set and extends over more
than 4 decades, and thus can be modeled with much higher
statistical accuracy than even the solar data sets
(Fig.~2). The size distribution of the KEPLER flare 
luminosities has a power law slope of $\alpha_F=1.813\pm0.004$,
which is similar to the hard X-ray fluxes of solar flares
(with $\alpha_F=1.79\pm0.06$, averaged from Fig.~(2a, 2b, 2c)).
It is gratifying to find similar power law slopes for solar 
and stellar flare fluxes, which can be interpreted in terms 
of an identical physical process operating in both cases 
(such as the magnetic reconnection process). 

The size distribution of stellar flare luminosities observed
by KEPLER is best fitted with the extreme event model (EM),
exhibiting a goodness-of-fit of $\chi=1.48$ (Fig.~4):
The inertial range $[x_0=7 \times 10^{33},
x_3=1.5 \times 10^{37}]$ erg is bracketing the power law
part, over which our model fits apply. 
Some exponential cutoff around $x_3=q_{exp} x_2
=0.04\ x_2 \approx 1.5 \times 10^{37}$ erg 
indicates finite system-size behavior, 
and the upper range range $[x_0, x_2]$ (with the maximum 
size value at $x_2=4 \times 10^{38}$ erg) indicates extreme 
events beyond the exponential cutoff, making up for
a fraction of $q_{pow}=0.11$. It appears that 11\% of
the flare sizes belong to a group of high-energy flare
events with extreme energies of $x \lapprox x_3$, while
the remaining 89\% of the flares constitute a different
group with low-energy events. 
  
\subsection{	Stellar Spectral Types  	}

Since the entire KEPLER flare catalog contains at least
6 different stellar spectral types (Davenport 2016;
Yang and Liu 2019), it is of interest
to study which spectral types produce the most extreme 
events and whether finite system-size effects occur
that give us some stellar diagnostics. For instance,
is the depth of the stellar convection zone related 
to the finite system-size inferred from size distributions?

We make use of the stellar spectral type classification
of the KEPLER stellar flare catalog (A, F, G, K, M, Giants),
for which the size distributions and power law fits are
presented in Fig.~3 of Yang and Liu (2009),
which is reproduced in Fig.~5 here. A list of
the power law slopes $\alpha$ obtained by Yang and Liu (2009),
with power law slopes $\alpha_P, \alpha_F, \alpha_E$
of our three models and the mean values (of the 3 models)
are provided in Table 2. The agreement of the power law
slopes $\alpha$ between the two studies is largely compatible: 
The differences are generally small,
14\% for F-type stars, 9\% for G-type, 1\% for K-type,
16\% for M-type, 5\% for Giants, except for A-type
stars (with a difference of 32\%). 
We attribute this discrepancy of the power law slope
in A-type stars ($\alpha=1.12\pm0.08$ according to
Yang and Liu (2009) versus $\alpha=1.65\pm0.07$
in Fig.~5 here) to a flawed fit by an inadequate
choice of the fitting range, which nullifies the
claim of Yang and Liu (2009) that A-type stellar 
flares are produced by a different physical mechanism. 
The model fits shown in Fig.~5 exhibit a straight
power law for K-type stars only (Fig.~5d), which shows
also the best agreement (1\%) between the two studies.
All other spectral types exhibit significant deviations
from straight power laws, which explains the discrepancies
between the power law fits of Yang and Liu (2009) and
our model fits. While arbitrary inertial ranges $[x_0, x_2]$
have been used in the past to fit a power law,
which is ill-defined in the absence of straight power laws,
we recommend to use the Pareto distribution model (PM)
(Eq.~3) to accomodate for incomplete sampling of small
events, or more adequately, the model with finite
system-size effects (FM; Eq.~4), and the model of extreme
events (EM; Eq.~5). The goodness-of-fit $\chi \lapprox 2$ 
of these models is found to be acceptable for all stellar
spectral types (Fig.~5). 

We show a synopsis of the flare energy size distributions
for the 6 spectral types of stellar flares, juxtaposed
with the size distributions of all stellar and solar
flare events in Fig.~6. This diagram demonstrates that
incomplete sampling occurs up to energies of 
$E \lapprox 10^{35}$ erg, while the power law-like
inertial ranges are confined to $E \approx 
10^{35}-10^{37}$ erg. The most energetic flare events
occur on Giant stars in the energy range of
$E \approx 10^{37}-10^{38}$ erg, perhaps capped by
extreme events at energies of  
$E \approx (1 - 5) \times 10^{38}$ erg.
This upper limit is slightly higher than derived
from sparse statistics of stellar flares
($E \lapprox 10^{37}$ erg) obtained before the KEPLER 
mission (see Fig.~3 in Schrijver et al.~2012).  

We find that the median or the maximum flare energy $E_{max}$ in
each stellar spectral type is related to the size of
the star, based on the mean or maximum of the stellar 
radius $R/R_{\odot}$ (Fig.~7),
defined by the Hertzsprung-Russell diagram 
of the Harvard spectral classification of main 
sequence and giant stars:

\par Giants, O-types $R \ge 6.6 R_{\odot}$,
\par B-types $R = 1.8-6.6 R_{\odot}$,
\par A-types $R = 1.4-1.8 R_{\odot}$,
\par F-types $R = 1.15-1.4 R_{\odot}$,
\par G-types $R = 0.96-1.15 R_{\odot}$,
\par K-types $R = 0.7-0.96 R_{\odot}$,
\par M-types $R \le 0.7 R_{\odot}$.

We show this relationship for the median 
flare energy $E_{med}$ (diamonds) in Fig.~7, 
as well as for the maximum flare energy $E_{max}$ 
(crosses in Fig.~7), sampled from 
the size distributions of the KEPLER data
(Fig.~6). This relationship supports the
notion that the maximum flare energy $E_{max}$ 
scales with the spatial size (or stellar
radius $R$), stellar surface area, or vertical
depth of the stellar convection zone. 
In Fig.~7 we show also the relationship
$E \propto (R^2 + R_M^2)$, which expresses
that the stellar area $A=\pi R^2$ is an upper 
limit of the flare energy for Giants and A-type
stars (since $R > R_M$), while M, K, F, G-type
stars have flare energies with a lower limit
at $R_M=0.7 R_{\odot}$ (since $R < R_M$), 
possibly indicating a lower limit of the
emission from the stellar corona, besides the 
chromospheric emission.
Consequently we can associate the largest
flare energies in the stellar flare size
distribution with the finite system-size
limit. 

\section{	Conclusions			}	

Occurrence frequency distributions (or size distributions)
come in all shapes and sizes. The most common case is the 
power law function, which is easily to recognize by
its straight line in a log(Number)-log(Size) representation,
but one finds curved functions also that exhibit significant
deviations from straight power laws and thus require 
refined models that can accomodate incomplete sampling,
instrumental sensitivity limitations, event-unrelated
background noise, truncation bias, finite system-size
effects, and outliers of extreme events, possibly caused
by multiple physical mechanisms, rather than a singular 
power law function produced by a classical SOC model.
In order to illustrate the meaning of finite 
system-size effects in the context of SOC models 
see also APPENDIX B and Fig.~10.
In this study we focus on the existence or absence of
finite system-size effects, which requires large
statistics with $n_{ev} \gapprox 10^4$ events per sample.
We apply our data analysis to solar and stellar data sets,
as well as to other empirical ``real-world'' data sets of
interest to our human population (Appendix B). 
Our conclusions are summarized in the following.

\begin{enumerate}

\item{We define 3 generalized power-law functions (Fig.~1):
(i) the Pareto distribution that shows a flattening at the 
lower end due to incomplete sampling, and a
straight power law at the upper end model (PM); 
(ii) the Pareto
distribution with an exponentially dropping cutoff at the
upper end that is caused by finite system-size effects (FM);
and (iii) an additional power law component that extends above
the exponential cutoff, caused by outlier events that are also called
``Dragon-King'' extreme events model (EM). 
We find that each of the
3 model functions can fit most of the observed 24 size distributions
analyzed here 
with a goodness-of-fit of $\chi \lapprox 2$. This quantitative result 
is superior to previous straight power law fits, where (i) arbitrary
(inertial) fitting ranges are used, (ii) power laws are ill-defined
for curved size distributions, and (iii) no goodness-of-fit is specified.}

\item{We re-visit solar flare size distributions of the hard X-ray peak
flux, the hard X-ray fluence, and flare durations (Figs.~2, 3). We corroborate
previous findings that the peak fluxes and the fluences can be well 
fitted by Pareto distributions, with no sign of finite system-size
effects at the upper end, indicating that larger super-flares are
possible above the currently observed maximum size limit of 
$L \lapprox 200$ Mm. For flare durations we find that the best fits
display an exponential cutoff at the upper end of $T_2 \lapprox
6$ hrs, which suggests that solar flares cannot last longer in a
coherent way. Size distributions of flare durations show no straight
power law part, partially because of relatively smaller inertial 
ranges.}

\item{The KEPLER mission provides the largest statistics 
($n_{ev} \approx 2.6 \times 10^5$) of stellar flare energy 
size distributions, and thus facilitates the most detailed
modeling of size distributions (Fig.~4). 
For the total of all KEPLER
observed flares we find an approximate inertial range of
$E \approx 10^{34}-10^{37}$ erg, a finite system-size range of
$E \approx 10^{37}-10^{38}$ erg, and extreme events at
$E \approx (1-5) \times 10^{38}$ erg.
For smaller subsets of KEPLER data (e.g., Fig.~5), the detection
of finite system-size effects is less feasible).}

\item{Subdividing the KEPLER flares into 6 stellar spectral types
(Figs.~5,6) we find that the maximum flare energy $E_{max}$ in
each stellar spectral type is related to the stellar radius
$R/R_{\odot}$ (Fig.~7), based on the mean stellar radius 
defined by the Hertzsprung-Russell diagram of the Harvard 
spectral classification. This relationship supports the
notion that the maximum flare energy may scale also
with the vertical depth $z$ of the stellar convection zone, 
based on a convection cell geometry with compatible vertical
and horizontal sizes, $z \approx 2R$. Consequently we
expect that the stellar surface area $(4\pi R^2)$ represents 
a finite system-size limit for flare energies in every 
stellar spectral type.}

\end{enumerate} 

This study demonstrates that we can extract a large
amount of new knowledge from model fitting of power 
law-like size distribution functions, sampled from a 
host of interdisciplinary data sets, pertaining to
nonlinear energy dissipation events in SOC models.
Although power law distributions are considered as
the hallmark of SOC models (Bak et al.~1987, 1988;
Aschwanden 2011; Aschwanden et al.~2016), a large
number of pertaining data sets exhibit significant
deviations from straight power laws, which justifies
the eloquent criticism of Stumpf and Porter (2012). 
We have demonstrated here that refined modifications
of power law-like functions, such as Pareto
distribution functions, exponential cutoff functions,
and dual (or multiple) power law functions are essential for
more accurate fitting of observed size distributions. 
Future work may focus on finite system-size effects 
in solar and stellar flares, which may provide new
knowledge on stellar convection and dynamos.

\section*{ APPENDIX A: Finite System-Size Effects in Empirical Data }

In the spirit of interdisciplinary research we fit our
power law-like size distribution functions to 8 empirical
data sets compiled by Clauset et al.~(2009). Four out of 
these 8 cases show mostly Pareto distributions with a
straight power law function all the way to the upper
end of the size distribution (Fig.~8), covering 
information technology (word frequencies, surnames, 
and weblinks), as well as geophysics 
(Earth quakes). Another four data sets (Fig.~9) 
exhibit clear deviations from straight power laws,
such as wildfires and human activities
(city sizes, blackouts, terrorism statistics).
Our size distribution models FM and EM indicate
some sort of finite system-size effects, either
in form of an exponential cutoff (see wildfires
in Fig.~9b), or in form of a secondary power law
component of extreme events beyond the cutoff
(see city sizes, blackouts, terrorism in Fig.~9).
The two power law components in the size distribution
may indicate some duality of scales: 
wildfires may be subdivided into natural 
(e.g., by lightning) and human-induced fires 
(Krenn and Hergarten 2009);
city sizes may be subdivided into two groups, 
depending on whether suburban agglomerations are 
merged with urban areas or not. 
The Zipf law does not fit as well with the data 
when using a traditional administrative definition 
of cities (Veneri 2013). 
Blackouts may be subdivided depending
on whether a detected blackout produces a
continent-wide ripple (cascading) or not
(Shuvro et al.~2018). 
Acts of terrorism may have different statistics,
depending on whether they are organized 
for economic causes, or for political causes (Kirk 1983).
The identification of such dual behavior could 
give us a deeper understaning of complex size
distribution functions that show significant 
deviations from ideal power laws. 

\section*{ APPENDIX B: Finite System-Size Effects in SOC }

All known SOC systems have a finite system size. For instance,
solar and stellar flares are limited by their finite 
solar or stellar surface area. Earth quakes or wild fires are 
ultimately bound by the Earth's continents, etc. In the original
sandpile model (Bak et al. 1987, 1988), upper limits on
avalanche sizes are given by the geometric size of
lattice 1-D, 2-D, or 3-D grids. The largest events span
the distance from the central region to the remote
boundaries of a sandpile. We illustrate the phenomenon
of finite-size effects in Fig.~10. In the fractal-diffusive
SOC model of a slowly-driven SOC system (Aschwanden 2012),
the centroid of an avalanche propagates according to a
random walk within the boundaries of a finite lattice grid
and stops at the edge of the lattice grid (thick solid
linestyle in Fig.~10), while the itinerary can cross the 
lattice boundaries, return to the inside area of the 
lattice, or even cross the boundaries multiple times 
(thin solid linestyle in Fig.~10), performing a random
walk pattern. The radial extent of the avalanche is also
called {\sl radius of gyration} (Charbonneau et al.~2001), 
which obviously can have
hugely different values for finite or infinite systems.
Numerical simulations of avalanches with finite system
size effects have been carried out in a number
of studies (Pruessner 2012), which mostly demonstrated
that the size distribution can be approximated with an
exponential cutoff function (Eq.~4). Alternatively,
some simulations show an additional bump near
the exponential cutoff (see p.~33 in Pruessner 2012).

\vskip1cm
{\sl Acknowledgements:}
We acknowledge helpful comments from the Editor
(Manolis Georgoulis) and an anonymous referee.
Part of the work was supported by NASA contract NNG04EA00C 
of the SDO/AIA instrument and NNG09FA40C of the IRIS 
instrument. 

\section*{ References }  

\def\ref#1{\par\noindent\hangindent1cm {#1}} 

\ref{Aschwanden, M.J. and Alexander, D. 2001,
        {\sl Flare Plasma Cooling from 30 MK down to 1 MK modeled 
	from Yohkoh, GOES, and TRACE observations during the 
	Bastille-Day Event (2000 July 14)}
 	SoPh 204, 91-121.}
\ref{Aschwanden, M.J. 2011
 	{\sl Self-Organized Criticality in Astrophysics. 
	The Statistics of Nonlinear Processes in the Universe},
 	ISBN 978-3-642-15000-5, Springer-Praxis: New York, 416p.}
\ref{Aschwanden, M.J. 2011,
	{\sl The state of self-organized criticality of the Sun
	during the last three solar cycles},
	SoPh 274:99-117.}
\ref{Aschwanden, M.J. 2012,
 	{\sl A statistical fractal-diffusive avalanche model of 
	a slowly-driven self-organized criticality system}
 	A\&A 539:A2.}
\ref{Aschwanden, M.J. and Freeland, S.M. 2012,
	{\sl Automated solar flare statistics in soft X-rays over
	37 years of GOES observations: The invariancxe of SOC
	during 3 solar cycles},
	ApJ 754:112.}
\ref{Aschwanden, M.J. 2015,
	{\sl Thresholded power law size distributions of instabilities
	in astrophysics}, 
	ApJ 814:19.}
\ref{Aschwanden, M.J. 2016,
	{\sl 25 Years of SOC: Solar and astrophysics},
	SSRv 198:47.}
\ref{Aschwanden, M.J. 2019,
	{\sl Self-organized criticality in solar and stellar flares:
	Are extreme events scale-free ?}
	ApJ 880, 105.}
\ref{Aschwanden, M.J. 2020,
	{\sl Global energetics of solar flares. XII. Energy 
	scaling laws},
	ApJ 903:23.}
\ref{Bak, P., Tang, C., and Wiesenfeld, K. 1987,
        {\sl Self-organized criticality - An explanation of 1/f noise},
        Physical Review Lett. 59/27, 381-384.}
\ref{Bak, P., Tang, C., and Wiesenfeld, K. 1988,
        {\sl Self-organized criticality},
        Physical Rev. A 38/1, 364-374.}
\ref{Charbonneau, P., McIntosh, S.W., Liu, H.L., and Bogdan,T.J.
 	2001, {\sl Avalanche models for solar flares},
 	SoPh 203, 321.}
\ref{Clauset, A., Shalizi, C.R., and Newman, M.E.J. 2009,
 	{\sl Power-law distributions in empirical data},
 	SIAM Review 51/4, 661-703.}
\ref{Borucki, W.J., Koch, D., Basri, G. et al. 2010,
	{\sl Kepler planet-detection mission: Introduction and
	first results} Science 327, 977.}
\ref{Davenport, J.R. 2016,
	{\sl The Kepler Catalog of stellar flares},
	ApJ 829:23.}
\ref{Drossel B. and Schwabl F. 1992,
	{\sl Self-organized critical forest-fire model.}
	Phys. Rev. Lett 69:1629-1632.}
\ref{Hergarten S. 2013,
	{\sl Wildfires and the Forest-Fire Model},
	in {\sl Self-Organized Criticality Systems}
	(ed., M.J. Aschwanden), Open Academic Press,
	www.openacademicpress.de}
\ref{Hill B.M. 1975, Annals of Statistics 3, 1163.}
\ref{Hosking, J.M.R. and Wallis, J.R. 1987,
	Technometrics 29, 339.}
\ref{Kirk, R.M. 1983,
	{\sl Political terrorism and the size of government: 
	A positive institutional analysis of violent 
	political activity}, Public Choice 40, 41–52, 
	https://doi.org/10.1007/BF00174995}
\ref{Krenn, R. and Hergarten, S. 2009,
	{\sl Cellular automaton modelling of 
	lightening-induced and man-made forest fires},
	Natural Hazards and Earth System Sciences 9:1743-1748.}
\ref{Lomax, K.S. 1954, 
	{\sl }, J. Am. Stat. Assoc. 49, 847.}
\ref{Malamud, B.D., Morein, G., and Turcotte, D.L. 1998, 
	Science 281:1840-1842.}
\ref{Pruessner, G. 2012,
 	{\sl Self-organised criticality. Theory, models and characterisation},
 	Cambridge University Press, Cambridge.}
\ref{Schrijver, C.J., Beer, J., Baltensperger, U. et al. 2012,
	{\sl Estimating the frequency of extremely energetic
	solar events, based on solar, stellar, lunar, and
	terrestrial records},
	JGRA 117, A8, CiteID A08103.}
\ref{Shuvro, R., Das, P., Wang, Z., Rahnamay-Naeini, M., 
	and Hayat, M. 2018. 
	{\sl Impact of Initial Stressor(s) on Cascading 
	Failures in Power Grid.} DOI 10.1109/NAPS.2018.8600585}
\ref{Sornette, D. 2009,
        {\sl Dragon-King, Black-Swans and the prediction of crises},
        J. Terraspace Science and Engineering, 2, 1}
\ref{Sornette, D. and Ouillon, G. 2012,
        {\sl Dragon-Kings: Mechanisms, statistical methods
        and empirical evidence},
        EPJST 205, 1}
\ref{Stumpf, M.P.H. and Porter, M.A. 2012, Science 335, 665.}
\ref{Taleb, N.N. (2007, 2010), {\sl The Black Swan: the impact 
	of the highly improbable} (2nd ed.). London: Penguin, 
	ISBN 978-0-14103459-1.}
\ref{Veneri P. 2013, 
	{\sl On City Size Distribution: Evidence from OECD 
	Functional Urban Areas}, OECD Regional Development 
	Working Papers 2013/27, 
	https://dx.doi.org/10.1787/5k3tt100wf7j-en}
\ref{Yang, H. and Liu, J. 2019,
	{\sl The flare catalog and the flare activity in the
	Kepler mission},
	ApJSS 241:1}
\ref{Zinck, R.D. and Grimm, V. 2008, 
	Open Ecol J 1:8-13, DOI 10.2174/1874213000801010008.} 


\begin{table}
\begin{center}
\normalsize
\caption{Observed phenomena with power law slopes, goodness-of-fit, and models
(PM = Pareto Distribution model, 
FM = Finite System-Size model, EM = Finite System-Size
and extreme event model).}
\begin{tabular}{lrrrrl}
\hline
Observable  	& Number & Background & Power law	& Goodness & Model  \\ 
		& of events & level   & slope 		& $\chi^2$ &        \\
\hline
\hline
Peak flux HXRBS	& 11,352 &          1 & $1.75\pm0.02$   & 0.88	   & EM \\ 
Peak flux BATSE &  7,245 &          1 & $1.76\pm0.02$	& 1.87	   & EM \\
Peak flux RHESSI&  7,998 & 	    1 &	$1.85\pm0.02$ 	& 1.41	   & PM \\
\hline
Counts HXRBS 	& 11,550 &          1 & $1.57\pm0.01$   & 1.52	   & PM \\ 
Counts BATSE 	&  3,425 &	    1 & $1.65\pm0.03$	& 0.90     & PM \\
Counts RHESSI 	& 11,549 &          1 &	$1.74\pm0.02$ 	& 1.51	   & PM \\
\hline
Duration HXRBS	& 11,549 &	    0 & $1.81\pm0.02$   & 1.81	   & EM \\
Duration BATSE	&  7,243 &	    0 & $1.99\pm0.02$   & 1.04     & FM  \\
Duration RHESSI & 11,525 & 	    0 &	$1.90\pm0.02$ 	& 1.40     & FM  \\
\hline
Kepler (catalog)& 162,264 &         1 & $1.813\pm0.002$	& 1.45     & EM \\
\hline
Words		&   9,695 &	    1 & $2.12\pm0.02$   & 1.85     & EM \\
Surnames	&   2,753 &         1 & $2.58\pm0.05$   & 1.72     & PM \\
Weblinks	&   8,658 &         1 & $1.49\pm0.02$   & 1.94     & PM \\
Earth quakes 	&  17,452 &         1 &	$1.71\pm0.01$   & 1.98     & EM \\ 
City sizes 	&  19,447 &         1 & $1.59\pm0.01$   & 1.57 	   & EM \\
Wildfires	&  56,052 &         1 & $1.77\pm0.01$	& 1.56     & FM  \\
Blackouts	&     213 &         1 & $1.35\pm0.10$   & 0.73	   & EM \\
Terrorisms 	&   4,303 &         1 & $2.64\pm0.04$   & 1.26	   & FM \\
\hline 
\end{tabular}
\end{center}
\end{table}

\begin{table}
\begin{center}
\normalsize
\caption{Model fitting of the power law slope of the size distributions
of stellar flares observed with KEPLER for three models 
(PM = Pareto Distribution model, FM = Finite System-Size model, 
EM = Finite System-Size
and extreme event model), separately sampled for six different stellar
spectral types (A, F, G, K, M, and Giants), and
compared with the previously published results 
of Yang and Liu (2009).}
\begin{tabular}{lrrrrrr}
\hline
Spectral & Number & Power law    & Power law  & Power law  & Power law  & Power law \\ 
type &of events &Yang and Liu (2009)& PM      & FM         & EM         & average  \\  		
	 &        & $\alpha$     & $\alpha_{PM}$ & $\alpha_{PM}$ & $\alpha_{EM}$ & $\alpha_{avg}$ \\
\hline
\hline
A-type	&    583 & $1.12\pm0.08$ & $1.65\pm0.07$ & $1.65\pm0.07$ & $1.65\pm0.07$ & $1.65\pm0.00$\\ 
F-type	&   8869 & $2.11\pm0.09$ & $1.96\pm0.02$ & $1.80\pm0.02$ & $1.79\pm0.02$ & $1.85\pm0.09$\\
G-type  & 55,259 & $1.96\pm0.04$ & $1.85\pm0.01$ & $1.80\pm0.01$ & $1.72\pm0.01$ & $1.79\pm0.06$\\
K-type  & 47,112 & $1.78\pm0.02$ & $1.69\pm0.01$ & $1.67\pm0.01$ & $2.03\pm0.01$ & $1.80\pm0.20$\\
M-type  & 50,439 & $2.13\pm0.05$ & $1.83\pm0.01$ & $1.83\pm0.01$ & $1.86\pm0.01$ & $1.84\pm0.02$\\
Giants  &  6,496 & $1.90\pm0.10$ & $2.39\pm0.03$ & $1.80\pm0.02$ & $1.80\pm0.02$ & $2.00\pm0.34$\\
\hline 
\end{tabular}
\end{center}
\end{table}


\begin{figure}
\centerline{\includegraphics[width=0.9\textwidth]{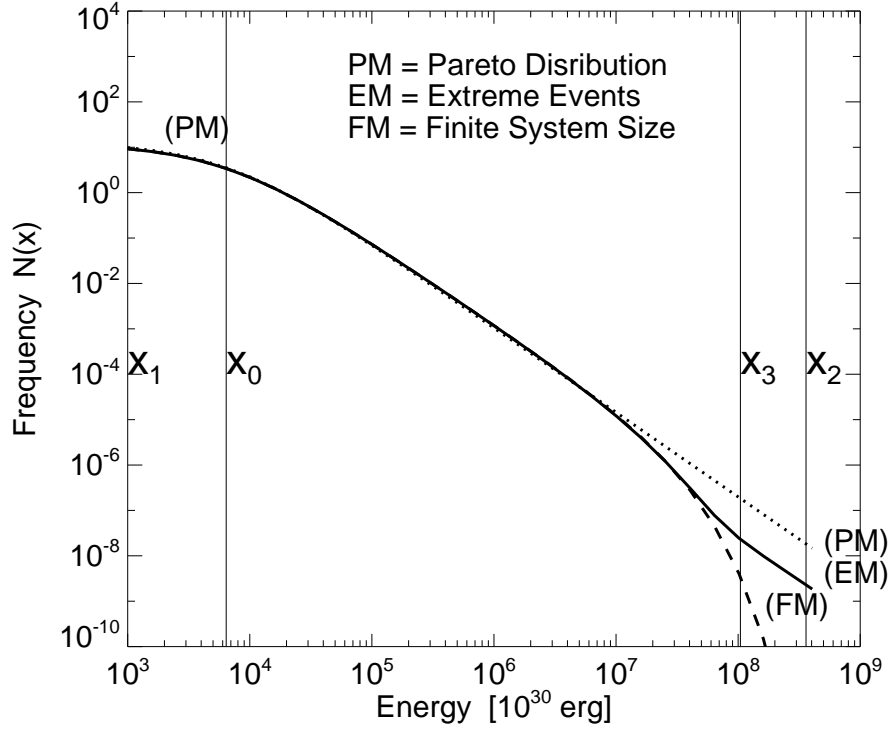}}
\caption{Synopsis of three power law models:
(PM) = Pareto distribution model $[x_1, x_0]$,
(EM) = Extreme events model $[x_3, x_2$], and
(FM) = Finite system-size model $[\gapprox x_3]$.
The inertial range covers $[\gapprox x_0, \lapprox x_3$].
The data used here are from the Kepler flare catalog,
see also Table 1 and Fig.~4.}
\end{figure}

\begin{figure}
\centerline{\includegraphics[width=0.9\textwidth]{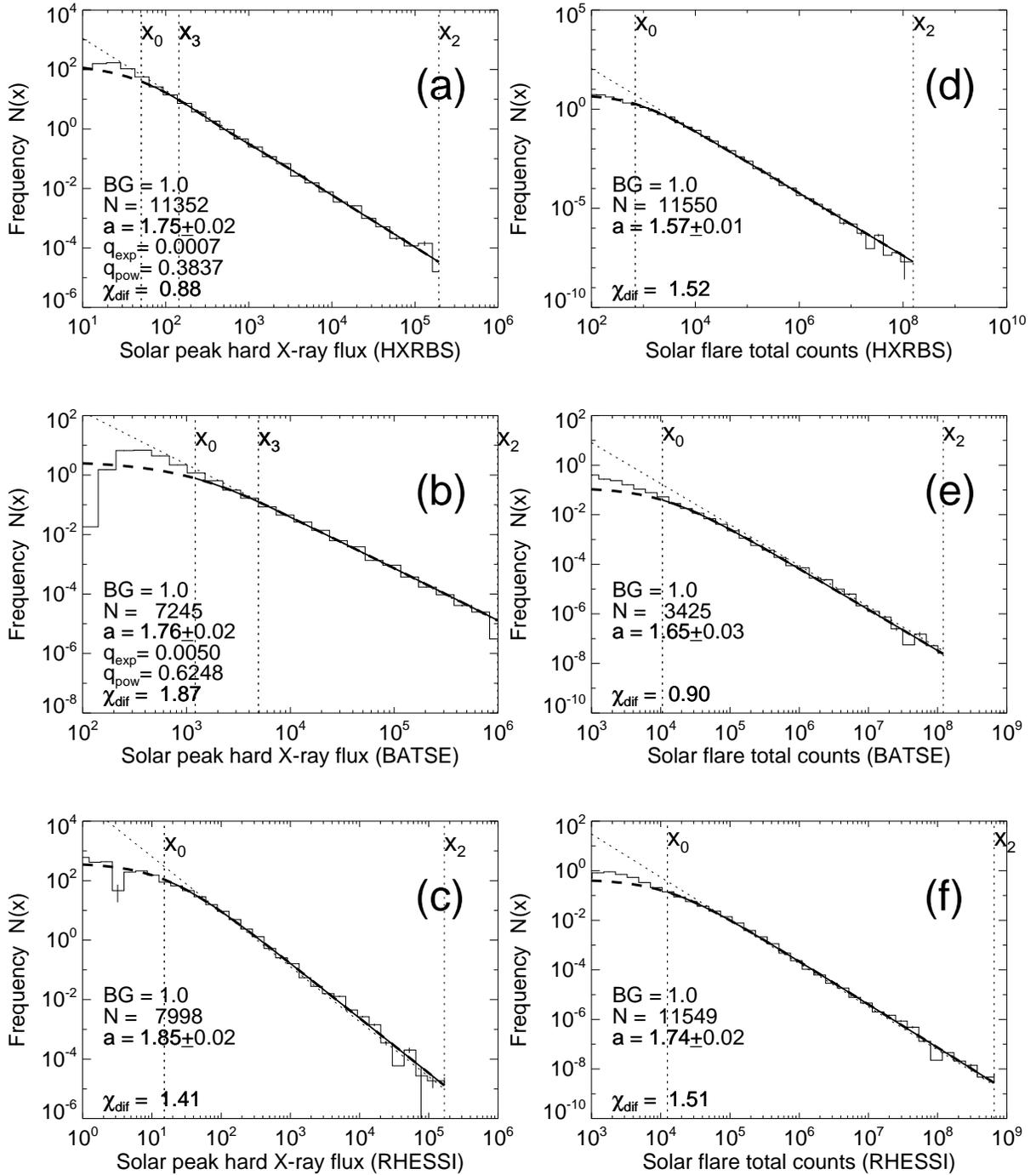}}
\caption{Differential occurrence size distributions of
solar hard X-ray peak fluxes (left panels) and total
counts (right panels), observed with three different
instruments (HXRBS, BATSE, RHESSI) (binned histograms)  
and fitted with the model functions (solid curves) over
an inertial range of $[x_0,x_2]$. For comparison, 
the power law slope is indicated with a dotted line.}
\end{figure}

\begin{figure}
\centerline{\includegraphics[width=0.9\textwidth]{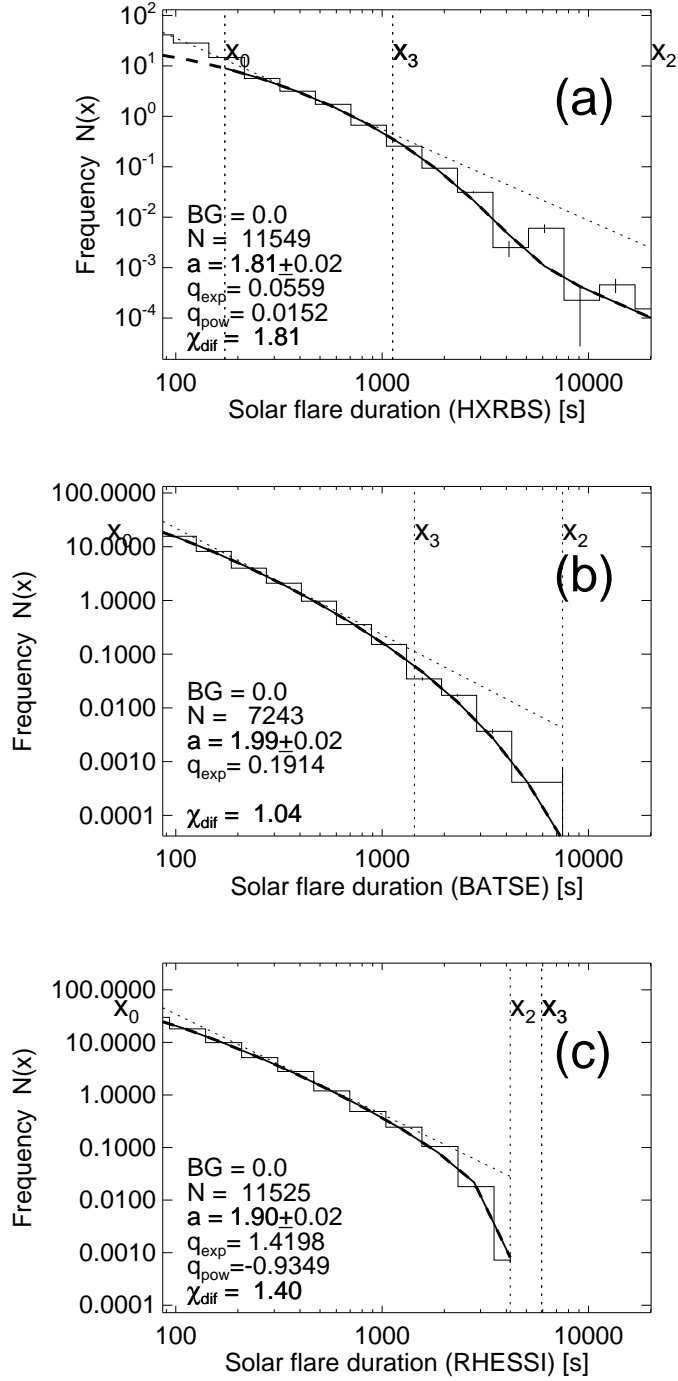}}
\caption{Differential occurrence size distributions of
solar flare durations, observed with three different
instruments (HXRBS, BATSE, RHESSI) (binned histograms)  
and fitted with the model functions (solid curves) 
over an inertial range of $[x_0,x_2]$. The size $x_3$ 
refers to the finite system-size model.}
\end{figure}

\begin{figure}
\centerline{\includegraphics[width=0.9\textwidth]{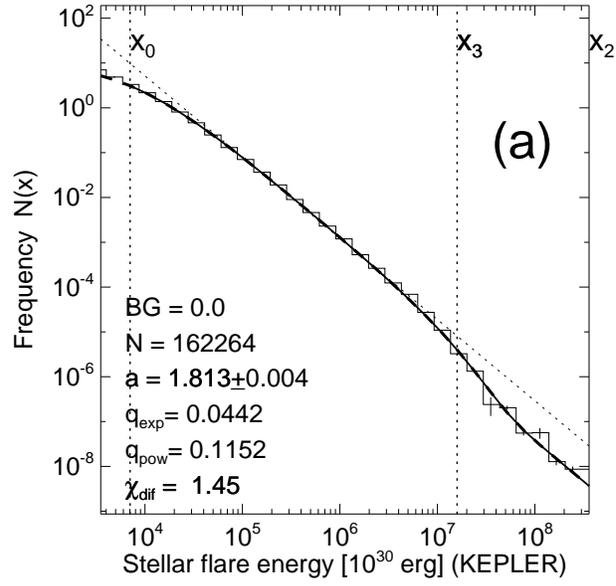}}
\caption{Differential occurrence size distributions of
stellar flare energy observed with KEPLER 
(binned histograms)  
and fitted with the model function (solid curve) 
over an inertial range of $[x_0,x_3]$. The size $x_3$
refers to the finite system-size model.
The power law slope is indicated with a dotted diagonal.}
\end{figure}

\begin{figure}
\centerline{\includegraphics[width=1.0\textwidth]{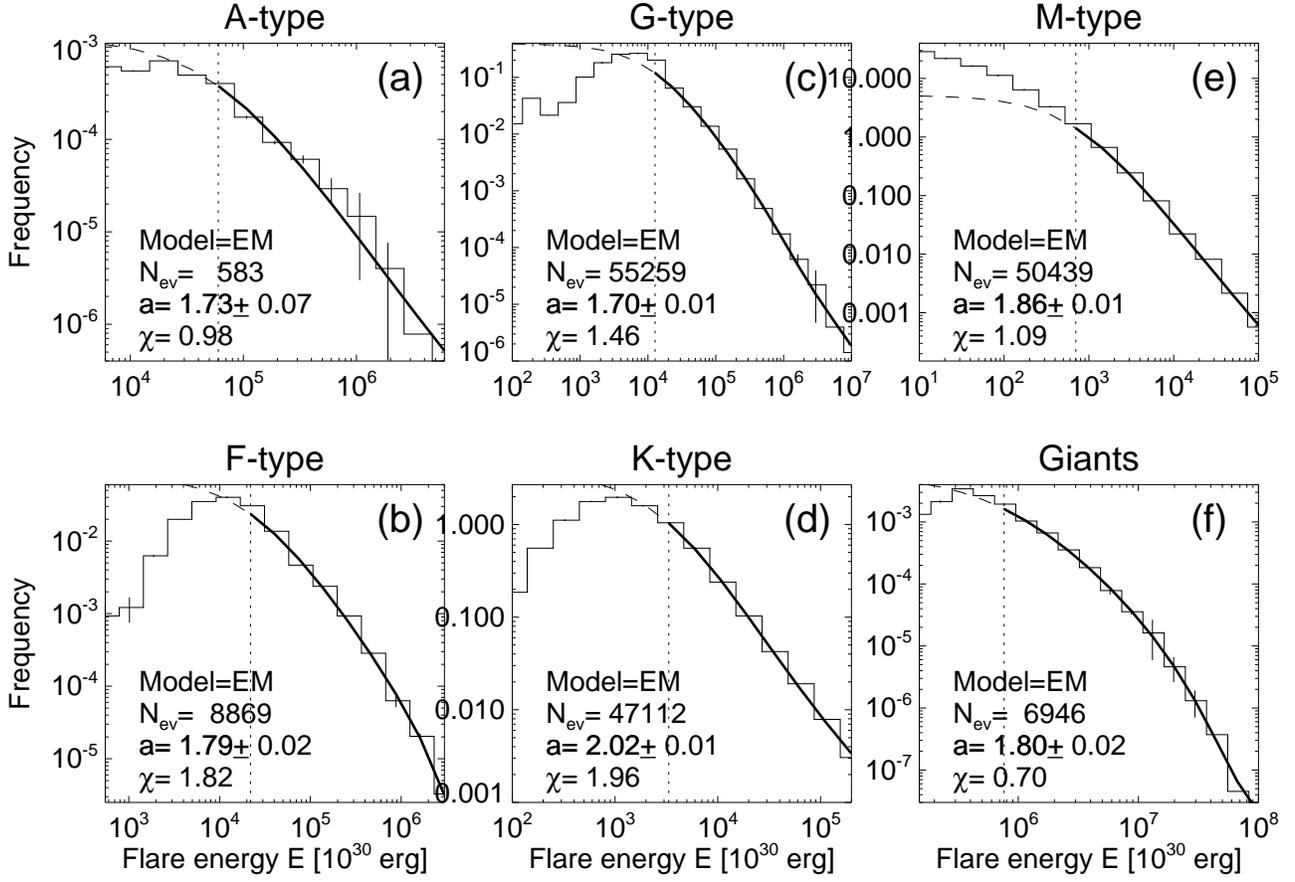}}
\caption{Size distributions of stellar flare energies 
observed with KEPLER in six groups of different spectral 
types (histograms), reproduced from Fig.~3 in Yang and Liu
(2019). The size distributions are fitted with our model
EM for extreme events (thick solid curves). Note the
deviations from straight power laws.} 
\end{figure}

\begin{figure}
\centerline{\includegraphics[width=1.0\textwidth]{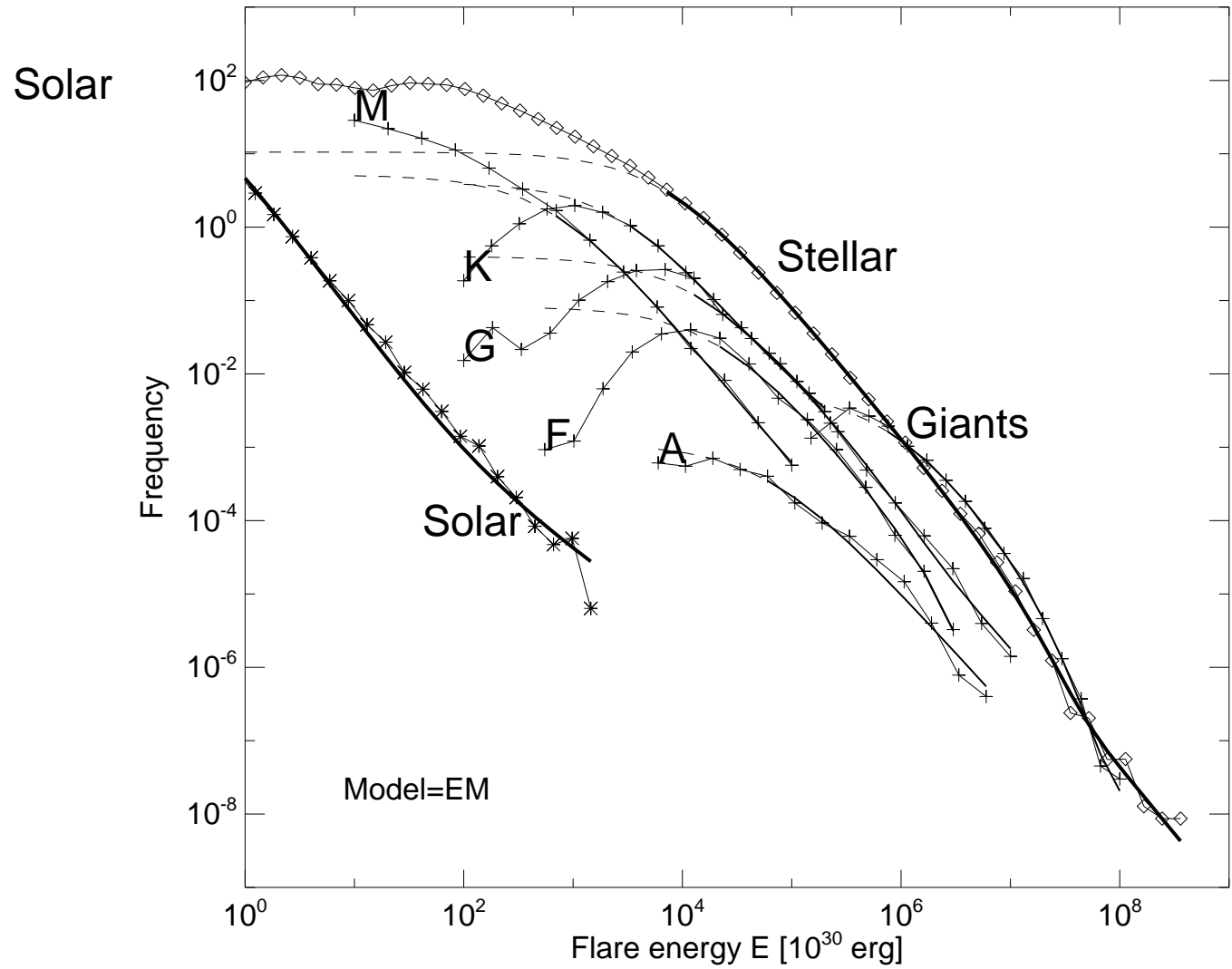}}
\caption{Composite size distribution of stellar flares (diamonds),
containing six groups with different stellar spectral 
types (A, F, G, K, M, Giants). For comparison we show
also the size distributions of solar (asterisks) and
all stellar flares (diamonds) combined (thick solid lines).}
\end{figure}

\begin{figure}
\centerline{\includegraphics[width=0.9\textwidth]{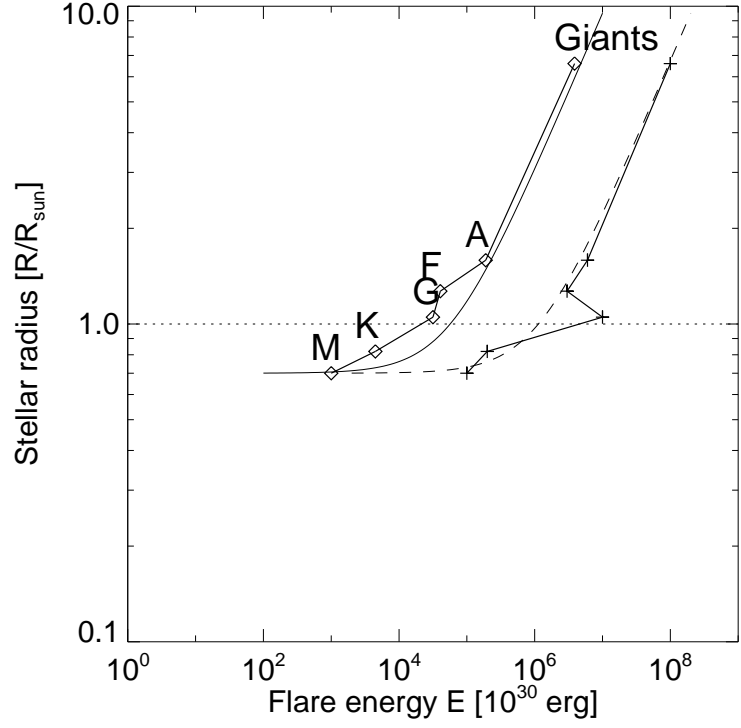}}
\caption{Observed relationship between the maximum
(curve with crosses) and median (curve with diamonds)
flare energies $E_{max}$ and the stellar radius in units of
solar radii, $R/R_{\odot}$, for 6 different stellar
spectral types. The thin curves depict the relationship
$E \propto (R^2 + R_M^2)$, where $R_M=0.7 R_{\odot}$ 
represents the lower limit of M dwarf stars.}
\end{figure}

\begin{figure}
\centerline{\includegraphics[width=0.9\textwidth]{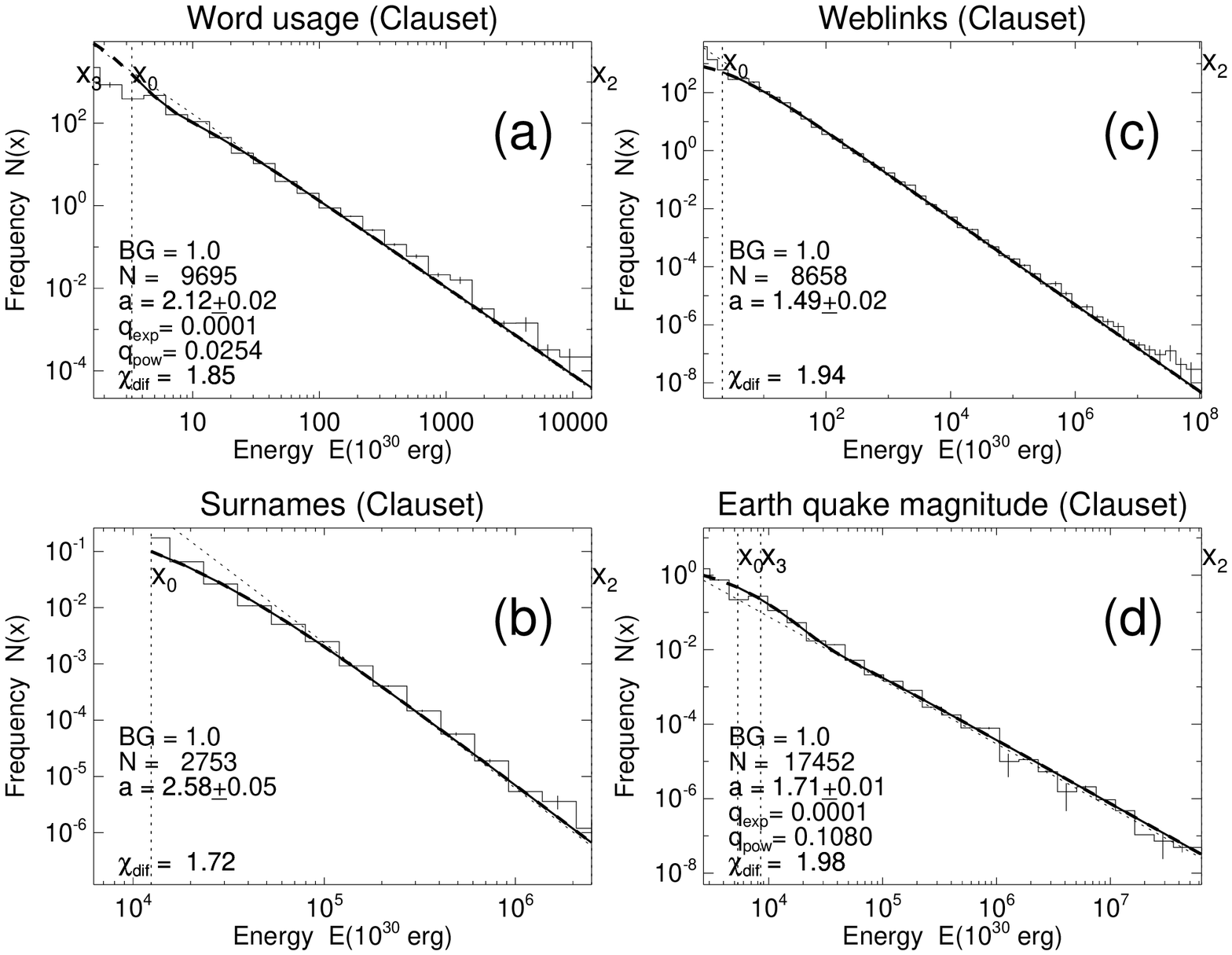}}
\caption{Differential occurrence size distributions of
empirical data sets compiled by Clauset et al.~(2009) 
(binned histograms) and fitted with the model functions 
(solid curves) over an inertial range of $[x_0,x_2]$.}
\end{figure}

\begin{figure}
\centerline{\includegraphics[width=0.9\textwidth]{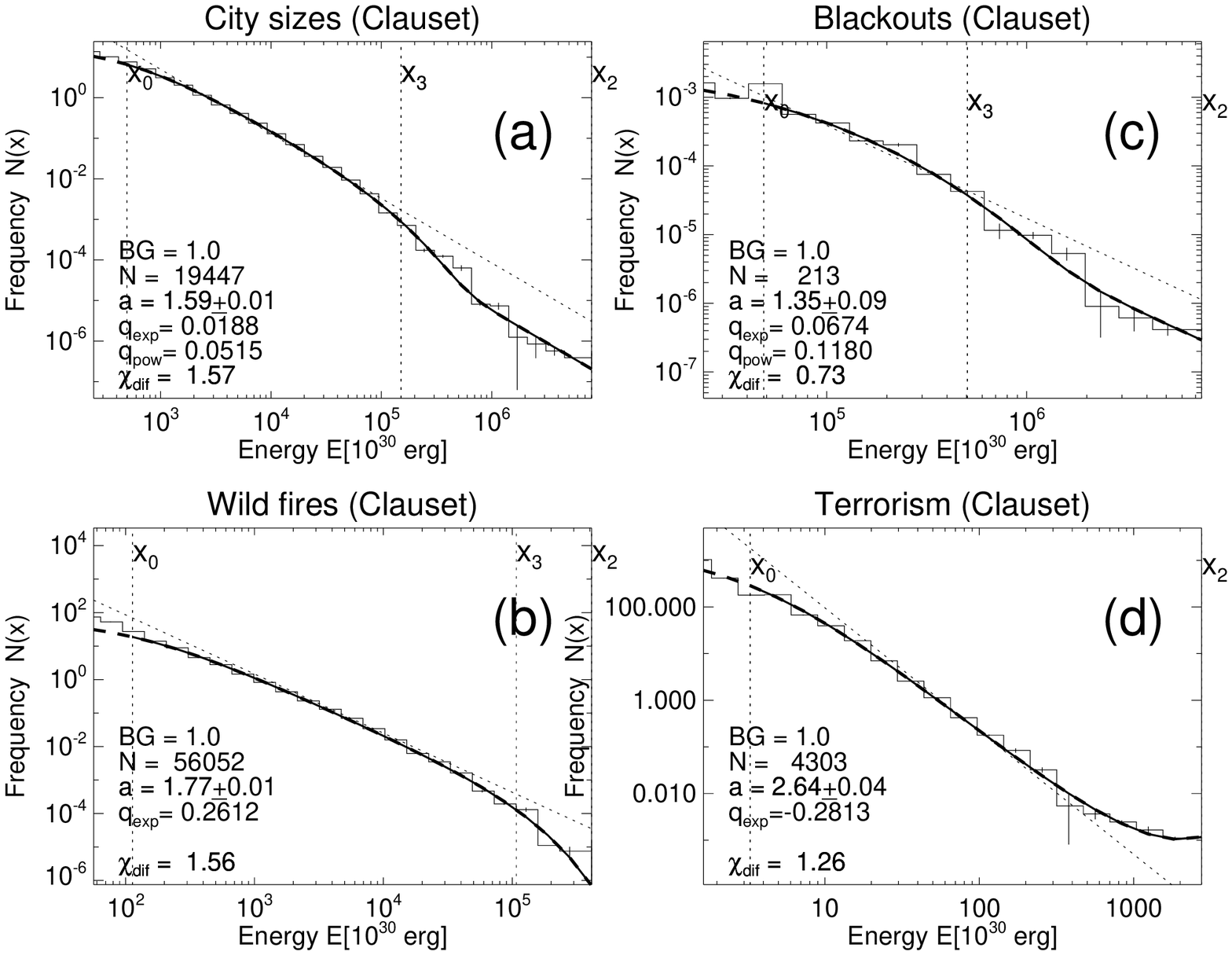}}
\caption{Differential occurrence size distributions of
empirical data compiled by Clauset et al.~(2009) 
(binned histograms) and fitted with the model functions
(solid curves) over an inertial range of $[x_0,x_2]$.}
\end{figure}

\begin{figure}
\centerline{\includegraphics[width=0.9\textwidth]{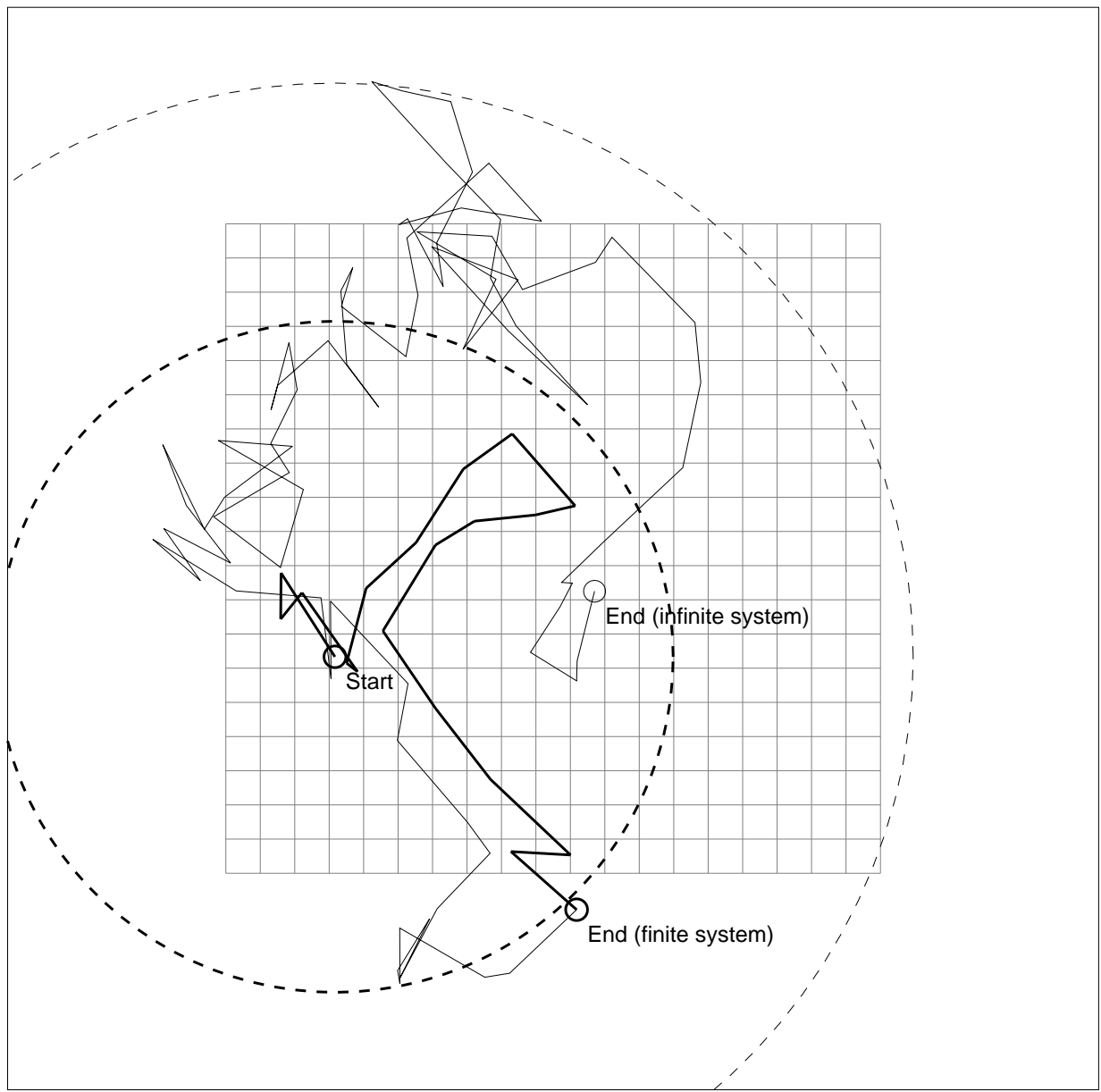}}
\caption{Example of 2-D random walk (or classical diffusion) in
a finite system size lattice grid (trajectory with thick
solid linestyle) and in an infinite system without
boundaries (trajectory with thin solid linestyle).
Both travel paths start at the same location, but end
at a different location, either at the system boundary 
or outside of the finite system. The gyration radii
of the two random walks are indicated with circles
in dashed linestyle, differing by a factor of about two.}
\end{figure}

\end{document}